# Insights from a Novel Tumor Model: Indications for a *Quantitative* Link between Tumor Growth and Invasion

Thomas S. Deisboeck [1,2,*], Yuri Mansury [1], Caterina Guiot [3,4], Piero Giorgio Degiorgis [4] and Pier Paolo Delsanto [4,5]

[1] Complex Biosystems Modeling Laboratory, Harvard-MIT (HST) Athinoula A. Martinos Center for Biomedical Imaging, HST-Biomedical Engineering Center, Massachusetts Institute of Technology, Cambridge, MA 02139 and [2] Molecular Neuro-Oncology Laboratory, Harvard Medical School, Massachusetts General Hospital, Charlestown, MA 02129, USA; [3] Dip. Neuroscience, Università di Torino, Italy and [4]INFM, Sezioni di Torino Università e Politecnico, Italy; [5] Dip Fisica, Politecnico di Torino, Italy;

**\*Corresponding Author:**

Thomas S. Deisboeck, M.D.
Complex Biosystems Modeling Laboratory
Harvard-MIT (HST) Athinoula A. Martinos Center for Biomedical Imaging
HST-Biomedical Engineering Center, Bldg. 16-352
Massachusetts Institute of Technology
77 Massachusetts Avenue
Cambridge, MA 02139
Tel: (617)-452-2226
Fax: (617)-253-2514
Email: <u>deisboec@helix.mgh.harvard.edu</u>

## ABSTRACT

Both the lack of nutrient supply and rising mechanical stress exerted by the microenvironment appear to be able to cause discrepancies between the actual, observed tumor mass and that predicted by West et al.'s universal growth model [*Nature,* **314**, 628 (2001)]. Using our previously developed model we demonstrate here, that (1) solid tumor growth and cell invasion are *linked*, not only qualitatively but also *quantitatively*, that (2) the *onset* of invasion marks the time point when the tumor's *cell density* exceeds a compaction maximum, and that (3) tumor cell invasion, reduction of mechanical confinement and angiogenesis can act *synergistically* to increase the actual tumor mass *m* towards the level $m_W$ predicted by West et al.'s universal growth curve. These novel insights have important implications for experimental and clinical cancer research alike.

# 1. INTRODUCTION

Volumetric growth and tissue invasion are hallmarks of highly malignant solid cancers. For multicellular brain tumor spheroids cultured *in vitro* within a three-dimensional gel environment, we already suggested a quantitative linkage between these key-features [1], yet to our knowledge experimental evidence at the *in vivo* scale is still missing. As a follow up work, we have recently proposed [2,3] a novel tumor growth model based on West et al.'s universal growth law [4]. Our numerical fitting results show that, using experimental spheroid data from the literature, malnourishment and confining mechanical stress can considerably reduce the actual tumor mass *m*.

Using this new mathematical model and based on the concept that there should be a maximum tumor cell density, $\rho_{Max}$, i.e. that there is a compaction threshold beyond which the cell density (i.e., the tumor cell mass per unit volume *v*) cannot be increased any further, we present in here the following arguments: (*a*) a possibility for tumor mass *m* to reach its asymptotic *M*, is by reducing the tumor cell density $\rho$ once its critical threshold, $\rho_{Max}$, is exceeded, (*b*) the biological equivalent for the reduction of cell density $\rho$ is the onset of cell invasion, linked therefore to the density threshold $\rho_{Max}$, (*c*) since we argue that $\rho$ and invasion are connected, and $\rho$ is linked to *v*, volumetric growth *and* invasion should be *linked* as well *through* $\rho$, (*d*) the *earlier* the onset of cell invasion, the *faster* the tumor's volumetric growth rate, and finally, (*e*) tumor cell invasion acts *synergistically* to increase nutrient supply and reduce mechanical resistance. Thus we combine here for the first time the cancerous key-features of growth *and* invasion, both qualitatively *and quantitatively,* and confirm the concept presented earlier in [1].

In the following we will briefly describe the relevant modeling steps together with the mathematical background for our argumentation.

## 2. THE MODEL

From the original West et al.'s law of energy conservation [4] it follows that, under ideal environmental conditions (i.e., fully replenished nutrient sources and in the absence of mechanical stress), the dynamics of tumor *mass* growth $m$ can be described by $dm/dt = am^p - bm$, where $p$ is a parameter whose value must be experimentally ascertained (West et al assume $p = 3/4$). The parameters $a$ and $b$ represent the rates of metabolism and nutrients intake, respectively [3]. Let $M$ represent the maximum body size attainable at the zero growth point $dm/dt = 0$, and define the auxiliary variable $y(t) \equiv 1 - [m(t)/M]^{1-p}$. The growth equation for tumor mass can then be further simplified as $dy/dt = -\alpha y$, where $\alpha = b(1-p)$. This first order differential equation has the closed-form solution $y(t) = \exp\{\ln[1-(m_0/M)^{1-p}] - bt\}$, which yields:

$$m_W(t) = M\{1 - [1 - (\frac{m_0}{M})^{1-p}] e^{-bt}\}^{1/(1-p)}, \qquad (1)$$

where $m_0$ denotes the initial tumor mass and $m_W(t)$ the value *predicted* by West et al.'s universal growth curve. Implicit in West et al's original model is the assumption that cell density $\rho$ stays constant throughout, $\rho(t) \equiv \overline{\rho}, \forall t$. However, as already argued in Delsanto et al. [3], to model the impact of increasing mechanical stress exerted by the microenvironment, it is more realistic to assume that proliferating solid tumors exhibit some ability to be *compacted*, if not compressed, under mechanical stress, so that the cell density $\rho$ *varies* over time. As a consequence, the variable of interest is no longer the tumor mass *m*, but the tumor *volume* $v(t) = m(t)/\rho(t)$. In that case, West et al.'s law of energy conservation can be restated as follows:

$$\rho \frac{dv}{dt} + v \frac{d\rho}{dt} = a(1-f)(\rho v)^p - b\rho v, \quad (2)$$

where the parameter $f \in [0,1]$ represents the reduction in the rate of tumor mass growth due to the compaction effect of non negligible, external mechanical stress. Hence, positive values of the parameter *f* imply a $m(t) - m_W(t)$ discrepancy between the actual tumor mass and the predicted level [3].

Let us now define *V* as the fixed-point value of the volume *v*, corresponding to the steady state point $dv/dt = 0$ under the above mentioned ideal conditions. We then introduce the maximum *threshold* level of tumor density, $\rho_{Max}$, defined by $d\rho/dt|_{\rho=\rho_{Max}} = 0$. Note that, while actual tumor density varies over time, $\rho = \rho(t)$, its threshold level $\rho_{Max}$ should remain constant throughout. Additionally, following [3], we define another auxiliary variable $z(t) \equiv 1 - [v(t)/V]^{1-p}$. With these definitions, the law of energy conservation becomes:

$$\frac{dz}{dt} = (1-z)(\alpha + \frac{dg/dt}{gR}) - \frac{\alpha(1-f)}{g}, \quad (3)$$

where $g(t) = [\rho(t)/\rho_{Max}]^{1-p}$. Equation (3) holds at all times $t$ except in equilibrium, where growth dynamics are assumed to return to its fixed point under ideal conditions: $a(\rho v)^p - b\rho v = 0$, i.e., without the growth inhibition factor $f$. As $t \to \infty$, assuming that tumor density converges into its steady-state value due to rising mechanical stresses, $\rho \to \rho_{Max}$ and $g \to 1$, Eq. (3) has the simple asymptotic solution $z = f$. Thus, as the system asymptotically approaches equilibrium at zero growth point, it can be readily verified that $v \to V(1-f)^{1/(1-p)}$. Note the distinction between the 'stree-free' equilibrium of West et al. with $f = 0$, and the asymptotic equilibrium with $f > 0$. The former essentially serves as the 'upper limit' of the latter beyond which tumor volume cannot be increased any further.

The above results have been derived *without* considering invasion explicitly. It is apparent however, that when the tumor cell density has reached its maximum threshold level, proliferative growth activity would come to a complete halt if invasion would not take place since further compaction to accommodate more daughter cells is no longer possible. Indeed, Eq. (2) implies that under rising pressure of compaction, as $f \to 1$, tumor growth becomes negative at the rate of $-b\rho v$, *unless* tumor cell invasion starts to take off and hence alleviates the pressures from mechanical confinements. In the following, let us therefore consider the case in which the onset of invasion is triggered when tumor cell density reaches its maximum threshold level.

Realistically, many solid malignant cancers consist of heterogeneous clonal sub-populations. As such, we first need to establish *geography* in the form of a spatially heterogeneous landscape, with the heterogeneity due to differential cellular density (e.g., as a consequence of different clonal cell proliferation rates). In modeling terms, such a "rugged" landscape is characterized by the interwoven pattern of "peaks" corresponding to maximum-density regions and "valleys" representing low-density areas. Next, suppose that the tumor landscape can be broken down into a countable collection of locations, *L*. The state of each location $\ell \in L$ depends on its onsite level of cellular density, $\rho_\ell \in [0, \rho_{Max}]$. Define the set *A* as the set of all locations exhibiting maximum density, $A \subset S = \{\ell : \rho_\ell = \rho_{Max}\}$.

Using computer visualization methods for data derived from a microscopic brain tumor growing within a three-dimensional biogel environment, Mojsilovic et al. [5] already described a *regional* onset of invasion from areas of high (proliferative) cell density. Based on this finding, we then introduce the critical assumption that the probability of cell invasion is monotonically linked to the onsite cell density level. In particular, the onset of invasion is triggered with high probability, $\text{Prob}_\ell(I)$, whenever the cell density reaches its maximum threshold level, $\text{Prob}_\ell(I) \to 1$ as $\rho_\ell \to \rho_{Max}$. This means that the set *A* contains all the maximum-density locations that serve as *launching pads* for tumor cell invasion. In this case, proliferative growth activity is not necessarily halted when cell density reaches its threshold level. Rather, invasion launched from those maximum-density locations allows tumor cells to migrate to more permissive areas with low levels of mechanical confinements, which in turn stimulates new proliferative growth activities.

With invasion, for the entire tumor system to converge to its asymptotic equilibrium *M*, eventually both the overall proliferative activity (i.e., the tumor's growth fraction) *and* cell invasion should experience a significant *slowing down*. Hence at equilibrium, we assume that the tumor's asymptotic cell density approaches a level that is strictly lower than the maximum threshold, i.e., $\rho_\ell \to \rho_\infty < \rho_{Max}, \forall \ell \in S$ as $t \to \infty$. A biological mechanism that could account for this behavior would be the volume expansion of edema fluid surrounding the main tumor mass, such as in the case of highly malignant brain tumors. While the tumor mass *m* is kept

Thomas S. Deisboeck et al. "Indications for a *Quantitative* Link between Tumor Growth and Invasion"

constant at it asymptotic value *M*, its volume *v* increases and thus the tumor density $\rho$ decreases, which in turn would slow down invasion.

To solve for the fixed point solution, set $dz/dt = 0 = dg/dt$ in Eq. (3). The asymptotic tumor volume, $v_\infty$, can then be easily derived for the case involving invasion as follows:

$$v_\infty = V[(1-f)/g_\infty]^{1/(1-p)}, \qquad (4)$$

where in equilibrium $g_\infty = \rho_\infty / \rho_{Max} < 1$, $\forall \ell \in S$. Otherwise, if regions with $\rho_\ell = \rho_{Max}$ exist in equilibrium, invasion is triggered again with $\text{Prob}_\ell(I)$ close to one and hence that stage cannot correspond to a fixed point. Therefore, due to the *critical link* between tumor growth *and* invasion, the asymptotic total tumor volume, $v_\infty$, will be significantly higher than in the no-invasion case: $v_\infty > V(1-f)^{1/(1-p)}$ precisely because $g_\infty < 1$. Thus, Eq. (4) allows for the possibility that, if an experimental (*in vitro* or *in vivo*) setup explicitly allows for invasion to occur, the actual tumor mass *m* may grow to the level predicted by the universal growth model of West et al.

## 3. DISCUSSION AND CONCLUSIONS

West et al.'s universal growth model [4] appears to be applicable not only to a variety of organisms but also to solid tumor growth [2]. In some instances, however, there is a discrepancy between the *actual*, observed tumor mass *m* and the level $m_W$ *predicted* by the universal growth curve. Using experimental *in vitro* data from the literature we recently found that, both malnourishment and the mechanical stress exerted by the tumor's microenvironment can contribute to a lower tumor mass *m* [3]. Hence conversely, both an increase in nutrient supply and a reduction in mechanical resistance should result in an upward shifting of the tumor growth curve, thus reducing the $m$-$m_W$ discrepancy [6].

Based on this previous work we argue in here further, that it is reasonable to assume that the extent of (interstitial) cell compaction is *finite*. As such, there should be a maximum cell density, $\rho_{Max}$, for any particular solid cancer type. Furthermore, since $\rho(t) = m(t)/v(t)$, any increase in *m* without a concomitant increase in *v* will lead to an increase in $\rho$. Concomitant or not, once $\rho_{Max}$ is reached, every additional increase in the tumor mass *m* as a result of proliferative activity would have to trigger an increase in *v* to keep the tumor cell density at a level $< \rho_{Max}$. One could therefore argue that maintaining cell density at a relatively low level, $\rho < \rho_{Max}$, is a driving force for rapid volumetric expansion of the tumor in a *pre*-invasive stage. However, within any confined three-dimensional microenvironment, one can expect that eventually there are limits for volumetric expansion. At this point, the options for the tumor system in order to sustain continuous growth are: (*a*) reducing the microenvironmental mechanical resistance and thus enabling continued volumetric growth; (*b*) reducing its own local cell density, $\rho$, through activation of cell motility into the adjacent tissue, and (*c*) *both*. Tissue invasion is mediated through the cells' secretion of matrix degrading enzymes, so called proteases, therefore is a disruptive process in nature. It is then reasonable to assume that cell invasion weakens tissue consistency adjacent to the main tumor, thus *reduces* the mechanical resistance and therefore contributes to a decreasing mechanical confinement for the proliferative tumor core. It is noteworthy, that in this concept angiogenesis would enable further volumetric growth and both, fractal vessel architecture as well as increased tissue edema (e.g., through a 'leaking' blood-brain barrier, in the case of malignant brain tumors) would likely contribute to a more sustained reduction in mechanical confinement.

We therefore propose that tumor cell invasion is triggered once the *maximum* cell density is reached. It then follows that since $v$ and $\rho$ are linked, and since we connect $\rho_{Max}$ to (the onset of) tumor cell invasion, volumetric tumor growth *and* cell invasion should be *linked* as well. We can thus advance the following hypothesis: in an experimental setup that allows for invasion to occur, i.e., in specific *in vitro* [1], *in vivo*, and clinical settings, we may observe the shifting *upwards* of the tumor-mass actual growth curve due to this mechanical pressure-reducing effect of cell invasion. The upward shifting of the tumor growth curve can elevate the tumor mass to $m_W$, i.e., the levels predicted by West et al.'s universal growth curve. Therefore, in an experimental or clinical setup that concomitantly examines both tumor cell proliferation and invasion we may observe a growth curve that fits the prediction of West et al.'s model more accurately.

One can further argue that within tissue types and areas of rather high mechanical confinement volumetric growth should quickly reach its limits and thus cell density must approach $\rho_{Max}$ fast. This in turn should trigger a comparatively *early* onset of invasion, arguably rendering the tumor *more* aggressive. It further follows that the *earlier* this onset of invasion is, the *faster* the resulting volumetric tumor growth rate will be *and* in addition, the faster will the tumor mass $m$ reach its asymptotic level $M$. Since this accelerated volumetric growth demands higher nutrient supply and since invasive cells reduce tissue resistance also for blood vessel formation, i.e. for approaching endothelial cells, it may trigger an early onset of *metastasis* as well, depending on the cancer type. We note that although the onset of invasion may at least in part be mechanically triggered (i.e., *passive*), this infiltrative process should be facilitated by an alternation in the cancer cells' gene-expression profile (e.g., producing proteases and extracellular matrix proteins) enabling an *active* phenotypic 'switch'. It is further noteworthy that we have introduced *geographical heterogeneity* in the form of distinct cell densities, which can be generated by variations in cellular proliferation rates due to the tumor's genetic instability. These cancerous subpopulations or clonal surface regions with high cell density may then function as "launching pads" for further tumor cell invasion, in turn enabling these regions to grow faster volumetrically than the low-density areas. As such, this recurrent "proliferation-migration-proliferation" sequence creates a closed *feedback loop*, which appears to favor the *competitive selection* of *more* aggressive (i.e. faster growing *and* faster invading) cell clones throughout the tumor surface. Thus, our theoretical framework here may also lead to novel insights into the process of *tumor progression*.

Our conclusions may therefore have *clinical* implications. Most importantly, Eq. (4) suggests that if numerical data consisting of both tumor volume (assessed through imaging) and cell density (biopsies) are available, it would be a straight-forward exercise to estimate the asymptotic mass of the tumor $M \approx v_\infty / \rho_\infty$, where $\rho_\infty < \rho_{Max}$ in equilibrium. Aside from such *predictive* purposes, our findings help to understand better some current *therapeutic* approaches and challenges. For example, anti-angiogenesis therapy which is geared towards limiting the tumor's blood supply would indeed result in a very desirable reduction of the tumor mass *m*. This is consistent with the experimental findings [7-10] showing that antiangiogenic therapy can effectively slow down or even inhibit tumor growth. Nonetheless, although a reduction in *m* vis-à-vis antiangiogenic treatment would also lead to a reduction in $\rho$ (unless it is accompanied by a comparable reduction in volume) and thus possibly reduce invasion temporarily, it however may facilitate rapid volumetric *recurrence* of the tumor as soon as the treatment stops. In contrast, reducing both *m* and *v* through surgical intervention does not (necessarily) lead to a reduction in $\rho$. It follows that, in case there is residual tumor left behind, cell invasion continues from this site and may contribute to overall treatment failure.

In summary, our model shows that (1) solid tumor growth and cell invasion are linked, not only qualitatively but also *quantitatively* as indicated by Eq. (4), thereby confirming the concept presented in [1], that (2) the *onset* of invasion marks the time point when the tumor's cell density exceeds the compaction maximum, and that (3) cell invasion, reduction of mechanical confinement and angiogenesis act *synergistically* in order to raise the actual tumor mass *m* towards the level predicted by the universal growth curve of West et al. [4] as well of the more general Delsanto et al.'s [3] version with a *generic* power-law exponent.

## ACKNOWLEDGEMENT

This work has been supported in part by NIH grant CA085139 (to TSD) and by the Harvard-MIT (HST) Athinoula A. Martinos Center for Biomedical Imaging and the Department of Radiology at Massachusetts General Hospital.